\documentstyle[12pt]{article}
\begin{document}
\def\lesssim{\stackrel{<}{\sim}}

\title{{\bf Timelike form factors at high energy}}

\author{Thierry Gousset$^{{\protect 1,2}}$ and Bernard
Pire$^{{\protect 2}}$}

\maketitle

\begin{center}
1. {\it DAPNIA/SPhN CE Saclay\\ 91191 Gif sur Yvette, France}\\
2. {\it CPT Ecole Polytechnique\\ 91128 Palaiseau, France}
\end{center}

\vspace{5cm}

\begin{abstract}
The difference between the timelike and
spacelike meson form factors is analysed in the framework of
perturbative QCD with Sudakov effects included. It is found that
integrable singularities appear but that the asymptotic behavior is
the same in the timelike and spacelike regions. The approach to
asymptotia is quite slow and a rather constant enhancement of the
timelike value is expected at measurable large \(Q^{2}\). This is
in agreement with the trend shown by experimental data.
\end{abstract}

\vspace{2cm}

\begin{flushright}
DAPNIA/SPhN 94 09 \\
hep-ph/9403293
\end{flushright}

\newpage

\section{Introduction}

\hspace{\parindent}
There is now a long history of continuous progress in the
understanding of electromagnetic form factors at large momentum
transfer. After the pioneering works~\cite{countrule} leading to
the celebrated quark counting rules, the understanding of hard
scattering exclusive processes has been
solidly founded by Brodsky and Lepage~\cite{brolep80}.

A perturbative QCD subprocess scaling like
\(\alpha_{S}(Q^{2})/Q^{2}\)
in the simplest case of the meson form factor is
factorized from a wave function-like distribution amplitude
\[\varphi(x,Q^{2})
=\int ^{Q} \psi(x,{\bf k}_{T})d{\bf k}_{T} \]
($x$ being the light cone fraction of momentum carried by
the valence quark), the \(Q^{2}\) dependence of which is analysed
in the renormalization group approach. Although an asymptotic
expression emerges from this analysis for the \(x\) dependence of
the distribution:
\[\varphi_{as}\propto x(1-x)\]
in the meson case, it was quickly understood that the evolution to
the asymptotic \(Q^{2}\) is very slow and that indeed some non
pertubative input is required to get reliable estimates of this
distribution amplitude at measurable \(Q^{2}\). Thanks to the QCD
sum rule approach, such a function was proposed by Chernyak and
Zhitnitsky~\cite{chezhi84}, which were followed by other model
dependent proposals~\cite{brafil89,farhul91}.

These developments helped theoretical estimates to get closer to
real experimental data but a severe criticism~\cite{isglle84}
remarked that most of the contributions to the form factor were
coming from end-point regions in the \(x\) integration, especially
when very asymetric distribution amplitudes such as those of
\cite{chezhi84} were used. This is not welcome since one
may doubt the validity of the perturbative calculation in these
regions. The recent work of Li and Sterman~\cite{li-ste92}
solves this problem by proposing a modified factorization formula
which takes into account Sudakov suppression of elastic scattering
for soft gluon exchange. This inclusion leads to an enlargement of
the domain of applicability of (improved) peturbative QCD
calculations of exclusive processes.

The case of timelike form factors has not been much studied
theoretically~\cite{hye---93,kropil93}. Experimental data on
the proton magnetic form factor \(G_{M}(Q^{2})\)~\cite{arm...93}
show a definite difference between the spacelike and timelike
values at the highest measured \(Q^{2}\). A recent analysis of the
\(\psi\rightarrow\pi\pi\) decay~\cite{kahmil93}
leads to a similar problem for the pion form factor at
\(O(10GeV^{2})\) transfer. Note, however, that the experimental
extraction of the spacelike form factor has been recently suspected
to suffer from large uncertainties~\cite{carmil90}.

In this paper, we carefully analyse in the Li-Sterman framework
\cite{li-ste92} the ratio between high \(Q^{2}\) timelike and
spacelike meson form factors. Not surprinsingly, we find that this
ratio goes asymptotically to \(1\) but we show that this approach to
asymptotia is slow and that factors of the order of \(2\) follow at
measurable \(Q^{2}\) from reasonable assumptions on wave functions.


\section{The spacelike form factor}
\label{the spacelike form factor}

\hspace{\parindent}
In this section, we review the formalism as it has been developed
for the spacelike case.

\subsection{Hard scattering picture}

\hspace{\parindent}
The spacelike form factor measures the ability of a pion to absorb
a virtual photon (carrying a momentum \(q\) with
\(q^{2}=-Q^{2}<0\)) while remaining intact. It is defined by the
formula:
\begin{equation}
\label{def FF}
<\pi(p')|J^{\mu}|\pi(p)>=e_{\pi}.F(Q^{2}).(p+p')^{\mu},
\end{equation}
where \(e_{\pi}\) is the pion electric charge and momenta are
defined in Fig.~\ref{sl-ff}.

In the hard scattering regime, that is when \(Q\) is very high
with respect to the low energy scales of the theory (the QCD
scale $\Lambda$ and the pion mass), Brodsky and Lepage have
motivated the following three step picture for the process,
valid in the light-front formalism~\cite{brolep79,brolep80}:
\begin{itemize}
\item
the pion exhibits a valence quark-antiquark {``}soft{''}
(see below) state,
\item
which interacts with the hard photon leading to another soft
state,
\item
which forms the final pion.
\end{itemize}

This leads to the convolution formula:
\begin{equation}
\label{convolution}
F(Q^{2})=\psi_{in}*T_{H}*\psi^{*}_{out}
\end{equation}
and the graphical representation of Fig.~\ref{conv}.

The most important feature of this picture is that it separates
hard from soft dynamics. The amplitude \(T_{H}\), the
{\em interaction}, reflects the hard transformation due to the
absorption of the photon and is hopefully calculable in
perturbative QCD, because the effective couplings are small in
this regime due to the asymptotic freedom. The amplitude $\psi$,
the {\em wave function}, which depends on low energy dynamics
is outside of the domain of applicability of perturbative QCD
and is, at present, far from being fully understood from the
theory. It is however process independent and contains much
information on confinement dynamics. Factorization proofs
legitimate this picture~\cite{CSSinMue}.


\subsection{Infrared corrections}

\hspace{\parindent}
The need of a careful factorization is due to the infrared
behavior of QCD: technically, large logarithms
(\(\sim\ln(Q/\lambda)\)) appear in the renormalized one-loop
corrections to naive {``}tree-graph{''} ($\lambda$ is some
infrared cut off needed to regularize soft and/or collinear
divergences). As in the renormalization procedure, if
factorization holds, these large corrections should be absorbed,
here in the re-definition of the wave function. The proof of
factorization and its consequences upon the wave functions are
studied in the pattern of the renormalization group. Without
entering into a detailed discussion, let us sketch the procedure
(see~\cite{botste89} for more on this leading
logarithms calculation and also for the renormalization group
treatment).

The first step is to compute the naive hard amplitude, that is
consider the tree graph of Fig.~\ref{tree-graph}
, and the three
other graphs related to it by C and T symmetries. One finds,
with notations explained on Fig.~\ref{tree-graph}:
\begin{equation}
\label{space-like}
T_{H}=16\pi\alpha_{S}C_{F}
{{xQ^{2}}\over{xQ^{2}+{\bf k}^{2}-i\varepsilon}}\hskip 0.265em
{{1}\over{xyQ^{2}+({\bf k}-{\bf l})^{2}-i\varepsilon}},
\end{equation}
where all quark momentum components are kept. Note
that we have done the usual projection onto the pion S wave
state: \(
\psi_{\pi}(p)\propto {{1}\over{\sqrt{2}}}\gamma^{5}p
\hskip -0.167em \hskip -0.167em \hskip -0.167em /\) and
used the C symmetry of the wave function.
\(C_{F}=4/3\) is the color factor, while \(
\alpha_{S}\) is the QCD effective coupling at the renormalization
point $\mu$.

To examine one loop corrections to \(T_{H}\), the relevant graphs
to consider in light-cone gauge are those of Fig.~\ref{1-loop}.

They directly lead to the wave function correction, in the
{``}double logarithms{''} or Sudakov region (namely:
\( \lambda \ll |{\bf q}| \ll u{{Q}\over{\sqrt{2}}} \ll
x{{Q}\over{\sqrt{2}}} \), $u$ and \({\bf q}\) being respectively
the light-cone fraction and transverse gluon momentum relatively
to the pion):
\samepage
\begin{eqnarray}
\psi^{(1)}(x,{\bf k})={{C_{F}} \over{2 \pi^{2}}}
\int _{\lambda}^{xQ/ \sqrt{2}}
{{d^{2}{\bf q}} \over{{\bf q}^{2}}} \alpha_{S}({\bf q}^{2})
\int _{|{\bf q}| \sqrt{2}/Q}^{x} {{du}\over{u}}
\{\psi^{(0)}(x-u,{\bf k}+{\bf q})
-\psi^{(0)}(x,{\bf k})\} \nonumber
\\
+{{C_{F}} \over{2 \pi^{2}}}
\int _{\lambda}^{\overline{x} Q/ \sqrt{2}}
{{d^{2}{\bf q}} \over{{\bf q}^{2}}} \alpha_{S}({\bf q}^{2})
\int _{|{\bf q}|\sqrt{2}/Q}^{\overline{x}}{{du}\over{u}}
\{ \psi^{(0)}(x+u,{\bf k}+{\bf q})
-\psi^{(0)}(x,{\bf k})\} ,
\label{one-loop}
\end{eqnarray}
where \( \overline{x} =1-x\) and the first term in the
difference comes from vertex-like corrections and the second
one
from self energy ones; in the infrared region some partial
cancellations occur between these corrections, but the
cancellation is not complete.

To pursue this analysis, it is convenient to define the Fourier
transform in the transverse plane:
\begin{equation}
\hat{\psi}{(x,{\bf b})}=\int d^{2}{\bf k}
e^{i{\bf kb}} \psi{(x,{\bf k})},
\end{equation}
and to separate transverse and longitudinal variations of the wave
function. One finds, omitting for the moment the second term in
Eq.~(\ref{one-loop}):
\begin{eqnarray}
\hat{\psi}^{(1)}(x,{\bf b})
={{C_{F}}\over{2\pi^{2}}}
\left(\int {{d^{2}{\bf q}}\over{{\bf q}^{2}}}
\alpha_{S}{({\bf q}^{2})}
(e^{-i{\bf qb}}-1)\int {{du}\over{u}}\right)
\hat{\psi}^{(0)}(x,{\bf b}) \nonumber
\\
+{{C_{F}}\over{2\pi^{2}}}
\int {{d^{2}{\bf q}}\over{{\bf q}^{2}}}\alpha_{S}({\bf q}^{2})
e^{-i{\bf qb}}
\int {{du}\over{u}}\left(\hat{\psi}^{(0)}(x-u,{\bf b})
-\hat{\psi}^{(0)}(x,{\bf b})\right).
\label{one-loop-b}
\end{eqnarray}

This equation contains the typical corrections one has to
consider in a hard process when dealing with either a big
(\(\gg 1/Q\)) or a small (\(\protect\lesssim 1/Q\)) neutral
object.
\pagebreak


\subsection{Transverse behavior at large distance}

\hspace{\parindent}
The transverse behavior at large distance is driven by the first
term of the previous equation, thanks to the vanishing of the
summation with the oscillating components. This occurs when
\(b={|{\bf b}|}\) is greater than at least a few times the inverse
of the upper bound of the corresponding integral:
\(xQ/\sqrt{2} \). As a consequence, in the remaining expression,
the infrared cut-off $\lambda$ can be replaced by the natural one
\(1/b\), above which the vextex and self energy corrections
do not compensate one another. Thus we get:
\begin{equation}
\hat{\psi}^{(1)}=-s(x,Q,b)\hskip 0.265em \hat{\psi}^{(0)},
\hskip 0.265em s={{C_{F}}\over{2\beta}}
\ln \hskip 0.265em {{xQ}\over{\sqrt{2}}}
\left( \ln {{\ln \hskip 0.212em xQ/\sqrt{2}}
\over{\ln \hskip 0.212em 1/b}}
-1+{{\ln \hskip 0.212em 1/b}\over{\ln \hskip 0.212em
xQ/\sqrt{2}}}\right),
\label{one-loop-large-b}
\end{equation}
with \(\beta=(11-{{2n_{f}}\over{3}})/4\), \(n_{f}\) being
the number of quark flavors. Here and in the following, it is
understood that the energies and inverse separations
are in the natural $\Lambda _{QCD}$ unit.

We have kept the single log term which occurs in the integration,
because it is the one necessary to express the true {\em dominant}
large \(b\) suppression, which one obtains in a more complete
treatment (that is leading and next to leading one)~\cite{li-ste92}.

After the ressummation of the ladder structure to all order, the
above Sudakov factor exponentiates. Taking into account the term
obtained with the substitution
\(x \rightarrow \overline{x} \equiv 1-x\), we get:
\begin{equation}
\hat{\psi} (x,b,Q)=e^{-s(x,b,Q)-s(\bar{x},b,Q)}
\hat{\psi} ^{(0)}(x,b,Q).
\end{equation}

Thus we get a strong suppression of the effective wave function as
\(b\rightarrow 1/\Lambda\), whatever the fraction $x$ is, provided
that $Q$ is reasonably large.

The remaining object \(\hat{\psi}^{(0)}\) is a soft component to
start with. It is soft in the sense that it does not include
loop-corrections harder than $1/b$. One may modelize it by including
some $b$ behavior or simply relate it to the distribution
amplitude~\cite{botste89} setting:
\begin{equation}
\hat{\psi}^{(0)}(x,b) \approx
\int _{0}^{1/b}\psi{(x,{\bf k})}d{\bf k}=\varphi(x;1/b).
\end{equation}


\subsection{Transverse behavior at small distances}

\hspace{\parindent}
The first term in Eq.~(\ref{one-loop-b}) is negligible when
the oscillating term remains close to $1$ in the range of
integration. This happens for $b$ a few times less than
\(\max ^{-1}(xQ,\overline{x}Q)\). In this case, soft divergences
cancel one another and one finds:
\begin{equation}
\hat{\psi}^{(1)}(x)
=\xi {{C_{F}}\over{2}}
\int _{0}^{1}dx' \left\{ {{\hat{\psi}^{(0)}(x')-
\hat{\psi}^{(0)}(x)}\over{x-x'}}
\theta (x-x')
+{{\hat{\psi}^{{(0)}}(x')-\hat{\psi}^{(0)}(x)}\over{x'-x}}
\theta (x'-x) \right\} ,
\label{one-loop-small-b}
\end{equation}
with the notation:
\begin{equation}
\xi={{1}\over{\pi^{2}}}
\int _{\lambda}^{Q}{{d^{2}{\bf q}}\over{{\bf q}^{2}}}
\alpha _{S}({\bf q}^{2})
\sim {{1}\over{\beta}} \ln \left({{\ln Q}\over{\ln \lambda}} \right).
\end{equation}

We displayed this equation in a slightly different form than in the
large $b$ case to explicitly show that Eq.~(\ref{one-loop}), in
the limit of small $b$, is related to the distribution evolution
proposed in~\cite{brolep79}.

Let us shortly review how this comes. Brodsky and Lepage had
proposed a simpler factorization formula for exclusive
processes~\cite{brolep80}. It is easily derived from the
previous treatment if one assumes that neither the wave function
nor the hard amplitude give important contribution to the
form~factor when the transverse momenta are big. Neglecting all
transverse momenta in $T_{H}$ leads therefore to consider the
$k_{T}$-integrated quantity:
\begin{equation}
\varphi(x)=\int d^{2}{\bf k} \psi (x,{\bf k})
\end{equation}

This distribution amplitude related to the wave function at $b=0$
has a dependence in $Q$ associated with the remaining
collinear  divergences
in Eq.~(\ref{one-loop-small-b}). Indeed, the exponentiated form of
this convolution equation, once it is written for the distribution
$\varphi$ and generalized to other regions than the Sudakov one,
leads to the celebrated expansion of
\(\varphi (x,Q)/x\bar{x}\) in a linear combination of a running
logarithm together with a Gegenbauer polynomial. However, whereas
this slow evolution is predictible, the expansion at some finite
$Q$ is inaccessible from perturbative reasoning.


\section{The timelike form factor}
\label{the timelike form factor}

\subsection{Quark and gluon poles}

\hspace{\parindent}
In the timelike region, the hard process ruling
\(\gamma^{*}\rightarrow \pi^{+}\pi^{-}\) is drawn in Fig.~\ref{tl-ff}
and the hard amplitude is simple to deduce from the spacelike
formula (\ref{space-like}) changing
\(p\rightarrow -p''\) or \(Q^{2}\rightarrow -W^{2},\hskip 0.265em
W^{2}=q^{2}\). The new feature with respect to the spacelike form
factor is that the contour of transverse momenta integration now
goes near poles located at either:
\({\bf k}^{2}=xW^{2}+i\varepsilon\) or:
\(({\bf k}-{\bf l})^{2}=xyW^{2}+i\varepsilon\).

These poles are automatically ignored in the pattern of the Brodsky
Lepage formalism as being to far from the contributing region of
integration. However, whereas this argument is reasonable at
asymptotic regime, we can expect some consequences of the presence
of these singularities when the energy is not so high.

Technically, these poles are, except in the end point regions
(\(x,y\rightarrow 0\)), far from the bounds of integration of the
two independant variables \(k={|{\bf k}|}\) and
\(K={|{\bf k}-{\bf l}|}\). Therefore, we may evaluate the integral
by deforming the contour of integration in the complex plane of
each of these variables.

Another question one may worry about, is the physical origin of
these poles. A complete physical amplitude, for example the
form~factor \(F{(Q^{2})}\), considered for complex value of
\(Q^{2}\), has poles and cut along the real negative axis
reflecting the existence of intermediate physical (on mass shell)
states. These intermediate states are hadronic ones and therefore
correspond to the {``}asymptotical{''} objects of confined QCD.
The poles we encounter in our present computation, internal gluon or
quark lines going on mass shell, of course, do not correspond to
observable states. However they only appear in a differential
amplitude which itself is not observable.\pagebreak Provided,
this differential amplitude is integrable, the resulting form factor
will only contain, as a remainder of this kind of singularities, a
real and
imaginary parts which one would also expect in a purely hadronic
computation.


\subsection{Hard scattering amplitude in b space}

\hspace{\parindent}
The Sudakov suppression is likely to be important at timelike
transfer, so it is necessary to take the transverse Fourier
transform of the hard amplitude. Furthermore, whenever this is
possible, it is interesting to get an expression not only for
negative $t$ or positive $s$, but also for complex values of
the generalized transfer. Let us define
\(z=\sqrt{-t}\hskip 0.265em ,\hskip 0.265em \arg (z) \in
\left[-{{\pi}\over{2}}\hskip 0.265em 0\right]\), so that in the
spacelike side of the complex plane, we get: $z=Q$ whereas in the
timelike side: $z=-iW$.

The expression of the form factor, with the Fourier transform of
Eq.~(\ref{space-like}) and the replacement of $Q$ by $z$, is:
\begin{eqnarray}
F=16\pi C_{F} \int dxdy
\int b_{1}db_{1}\hat{\psi} (x,b_{1})
b_{2}db_{2} \hat{\psi} (y,b_{2}) \hskip 0.265em
\alpha _{S} \hskip 0.265em T(b_{1},b_{2},x,y,z), \nonumber
\\
T=K_{0}(\sqrt{xy} zb_{2}) \hskip 0.265em xz^{2}
\left\{ \theta(b_{1}-b_{2})I_{0}(\sqrt{x} zb_{2})
K_{0}(\sqrt{x} zb_{1})
+(b_{2} \leftrightarrow b_{1}) \right\} ,
\label{timelike}
\end{eqnarray}
where angular integrations have been done thanks to the cylindrical
symmetry of both hard amplitude and S wave wave function.
$b_{2}$ and $|b_{1}-b_{2}|$ are the transverse distances of,
respectively, the gluon vertex and the internal quark vertex. The
functions $K_{0}$ and $I_{0}$ are modified Bessel functions of
order 0, the first appearing in the equation comes from the gluon
propagator, while the remainder comes from the quark propagator.


\subsection{Asymptotic behavior}
\label{asymptotic behavior}
\hspace{\parindent}
Bessel functions have different asymptotic behaviors in various
directions of the complex plane: for
\(|\zeta| \rightarrow \infty ,\hskip 0.265em
\arg (\zeta) \in \left[ -{{\pi}\over{2}} \hskip 0.265em
0 \right] \) we have~\cite{abramowitz},
\begin{equation}
K_{0}(\zeta)\approx \sqrt{{{\pi}\over{2\zeta}}}e^{-\zeta},
\hskip 0.265em I_{0}(\zeta) \approx \sqrt{{{2}\over{%
i\pi\zeta}}} \cosh (\zeta+i{{\pi}\over{4}}),
\end{equation}
we thus have to study how the asymptotic dependence of the form
factor is affected by this direction. In particular in the timelike
limit, the integrand is no more exponentially suppressed.

There is, {\it a priori}, no general constraint to ensure that
the limit of some observable like a form factor should be the same
in every directions in the complex plane. Even though \(F(z)\) is
analytic and:
\begin{equation}
F(z) \propto {{\alpha_{S}(z^{2})}\over{z^{2}}}(1+\varepsilon(z)),
\end{equation}
with the limit: \(\varepsilon(z)
\rightarrow ^{|z| \rightarrow +\infty }_{\arg z=0}0\), our
ignorance of the true form of $\varepsilon$ prevents us from
concluding when another direction is considered. However, because
one expects that the same kind of physics underlies exclusive
processes,  we expect that, at least in an asymptotic regime, we
should find, for the leading behavior, the
overlapping of exactly the same soft and hard amplitudes.

As a first step in the understanding of the features of the whole
form factor, we may evaluate the $b$ integrals in an analytical way,
putting a simple form for the wave function:
\begin{equation}
\hat{\psi}(b)=\theta(B-b),
\end{equation}
which automatically provides a cut-off to avoid indefinite
integral. We also forget here the possible running of the coupling
with transverse distances (see subsection~\ref{comparison}) which
appears with Sudakov suppression, here ignored.

With these simplifications, one finds for the integral over
$b_{1}$ and $b_{2}$:

\begin{eqnarray}
\label{asymptotic}
I=\int b_{1}db_{1} \hat{\psi}(b_{1})b_{2}db_{2} \hat{\psi}(b_{2})
T(b_{1},b_{2},x,y,z) \nonumber
\\
I={{1}\over{xy \hskip 0.265em z^{2}}}+{{\sqrt{B}}\over{z^{3/2}}}
f(x,y,zB),
\end{eqnarray}
with $f$ a function that we refrain from quoting here due to
its lack of interest, except for its generic behavior for large
\(xy|z|B\): \(f\sim e^{-zB}\).

As long as we avoid the timelike limit, we find the following
leading behavior for the integration in the transverse plane:
\begin{equation}
I \sim {{1}\over{xy\hskip 0.265em z^{2}}},
\end{equation}
which displays the expected selection of small configuration by the
hard process.

In the timelike region, this is no more true, as we get a modified
power dependence:
\begin{equation}
I\sim {{1}\over{W^{3/2}}}\sqrt{B} \hskip 0.265em e^{iWB},
\end{equation}
with the appearance of the size $B$, together with an oscillating
factor. Of course, we
suspect here that we have found essentially the limit of our model
object; nevertheless, we may anticipate that some reminiscence of
this
rather different behavior will occur in the non asymptotic regime.

A source of modification to the above result is the transverse
behavior of the wave function. The rectangular form which we have
choosen above and its steep variation reduces the occurence of
cancellations expected with an oscillating integrand. If we were
speaking of Fourier-transform we would say that a rectangular
function has relatively large components at large momenta in
comparison with any similar but smoother function.

The Sudakov behavior reviewed in the second section plays this
role. However, due to the transfer dependence of the Sudakov factor,
we must firstly face the problem of its analytic
continuation~\cite{magste90}.

Before turning to this analytic continuation, we rewrite the
expression of $s$ from Eq.~(\ref{one-loop-large-b}) for spacelike
transfer, in the form:
\begin{equation}
\label{s(x,Q,b)}
s(x,Q,b)={{C_{F}}\over{2\beta}} \ln
{{xQ}\over{\sqrt{2}}} \hskip 0.265em
{(U-1-ln\hskip 0.265em U)},\hskip 0.265em
U={{-ln\hskip 0.265em b}\over{ln\hskip 0.265em
{{xQ}\over{\sqrt{2}}}}},
\end{equation}
\pagebreak
to explicitly see that $s$ increases rapidly with
\(U < 1\) at large $Q$ (remember that if $x$ is small, we can
always consider the $\bar{x}$-term in turn) so that the region of
not too large suppression is \(U \protect\lesssim 1\), where we have:
\begin{equation}
s \approx {{C_{F}}\over{4\beta}} \ln
{{xQ}\over{\sqrt{2}}} \hskip 0.265em
(U-1)^{2}.
\end{equation}

For large timelike transfer, setting $Q=-iW$ in Eq.~(\ref{s(x,Q,b)}),
we get:

\begin{equation}
s(x,W,b)-s_{SL}(x,W,b) \approx i{{C_{F} \pi} \over{4 \beta}}
\ln U-{{C_{F} \pi^{2}} \over{16 \beta \hskip 0.265em
\ln \hskip 0.265em {{xW} \over{\sqrt{2}}}}}
\label{s-sSL}
\end{equation}
with $s_{SL}$ and $U$ the spacelike expression of
Eq.~(\ref{s(x,Q,b)}). In the above equation, the real part is
effectively small, while the imaginary part remains close to
$0$ in the region of intermediate suppression. We can
therefore assume, in the asymptotic regime, the same scale
dependence for the Sudakov factor and use the expression
$s_{SL}$ for our study.

To simplify our purpose, we will concentrate on the simpler case
one gets by considering only the transverse behavior of the gluon
propagator, that is setting the transverse momentum to
\({\bf k}={\bf 0}\) in Eq.~(\ref{space-like}). As we will show,
this does not alter the naive behavior we previously get. However
when taking the Fourier transform, only one transverse distance
remains, \({\bf b}={\bf b}_{1}={\bf b}_{2}\) and after angular
integration we are led to replace the integral I in
Eq.~(\ref{asymptotic}) by the quantity $I'$:
\begin{equation}
I'=\int _{0}^{\Lambda^{-1}}bdbK_{0}({{1}
\over{2}}zb)e^{-4s(|z|,b)}
\end{equation}
where we have limited our study to the \(x=y={{1}\over{2}}\) case.
Thanks to the Sudakov suppression, the upper
bound of the integral is naturally \(b=\Lambda^{-1}\).

For the Sudakov exponent, we consider the approximate expression of
section~\ref{the spacelike form factor} with the prescription of Li
and Sterman~\cite{li-ste92}
which is to set the exponential to unity in the region that should
not be controled by Sudakov evolution, here for \(b < 1/|z|\).
With these simplifications, we get:
\begin{equation}
I'={{4}\over{W^{2}}} \left( -1-iK_{1}(-i)
+\int _{1}^{W/2}uduK_{0}(-iu)e^{-4s(W,{{2u}\over{W}})} \right)
\end{equation}
which is to be compared with the expression without Sudakov
correction:
\begin{equation}
I={{4}\over{W^{2}}}(-1-i{{W}\over{2}}K_{1}(-iW/2))
\end{equation}

We present in Fig.~\ref{as} the result of a numerical computation
for both the real part (a) and the modulus (b) of the quantity
\begin{equation}
\Delta={{W^{2}}\over{4}}I+1
\end{equation}
which dictates the deviation from the counting rule canonical
result and compare it to the original quantity
\({{W^{2}}\over{4}}I+1\). We observe that after an intermediate
regime (\(W<20\Lambda\)), the expressions including Sudakov
suppression slowly decrease to $0$ contrary to the case of the
rectangular wave function.

Another feature we have omited until now is the fact that the
expression for the form factor is a superposition of amplitude
with various fractions $x,y$ weighted by smooth distributions.
This should also modify the $W^{-3/2}$-power law and we examine
this possibility in Appendix A.

We conclude that the $W^{-3/2}$ behavior does not resist to the
inclusion of any realistic $b$ or $x$-$y$ integration procedure.
Indeed the dimensional counting rules are valid at timelike
transfers and furthermore the form factors are asymptotically the
same.


\subsection{Comparison at intermediate energy}
\label{comparison}

\hspace{\parindent}
Let us now turn to the discussion of the ratio of timelike over
spacelike form factors in the intermediate range. The complete
integration formula is:
\begin{equation}
\label{F expression}
F=\int dxdy\int b_{1}db_{1}b_{2}db_{2}\hat{\psi}^{(0)}(x,b_{1})
\hat{\psi}^{(0)}(y,b_{2}) \hskip 0.265em T_{H} \hskip 0.265em
e^{-S}
\end{equation}
with the hard scattering amplitude from Eq.~(\ref{timelike}):
\(T_{H} = 16 \pi \alpha _{S} T\). The
integration range for the longitudinal fraction of momentum goes
from $0$ to $1$, whereas transverse distances it go from
$0$ to $1/\Lambda$ thanks to the Sudakov suppression at
large $b$.

For the numerical study, we take into account the one loop running
of the QCD coupling in the hard scattering:
\begin{equation}
\alpha_{S}(t)={{\pi}\over{\beta \hskip 0.265em \ln
(t^{2}/\Lambda^{2})}}, \hskip 0.265em t=\max
(\sqrt{xy}|z|,b^{-1}_{1},b^{-1}_{2})
\end{equation}
with the prescription for the renormalization point described
in~\cite{li----93} and \(\Lambda=200MeV\). The Sudakov factor
\(e^{-S}\) contains the corrections for the two wave functions
together with the anomalous running of the four quark operator
\(T_{H}\)~\cite{li----93}.

The Sudakov factor should be analytically continued as
discussed in sub-section~\ref{asymptotic behavior}~\cite{magste90}.
It turns out that this procedure leads to
quite model-dependent results at non asymptotic transfers. This has
to do with the truncation of the exponentiated expression and with
the need
to suspect the validity of the approach when the (real part of the)
Sudakov exponent becomes positive, i.e. when Sudakov suppression
turns to an enhancement. We leave to Appendix B a somewhat detailed
discussion of these effects, the conclusion being that an extra
modification of the timelike value may come from this continuated
Sudakov exponential, but that it is quite difficult to reliably
quantify this statement. In the following, we will thus keep the
Sudakov factor at its spacelike (real) value.

We used various forms for the soft wave function
\(\hat{\psi}^{(0)}(x,b)\) to test the sensitivity of the
result to this input. One may consider wave functions without
intrisic transverse behavior (\(\hat{\psi}^{(0)}(x,b)=\varphi(x)\))
and use either the asymptotic form:
\begin{equation}
\label{phi-as}
\varphi_{as}(x)=\sqrt{{{3}\over{2}}}f_{\pi}x(1-x),
\end{equation}
\pagebreak
with \(f_{\pi}=133MeV\) the pion decay constant, the CZ form:
\begin{equation}
\label{phi-CZ}
\varphi_{CZ}(x)=5(1-2x)^{2} \hskip 0.265em
\varphi_{as}(x),
\end{equation}
or other expansions in terms of Gegenbauer polynomials like those
proposed in~\cite{brafil89,farhul91}. In the
following, we will show some results for one form~\cite{farhul91}:
\begin{equation}
\label{phi-FHZ}
{
\varphi_{FHZ}(x)=(.6016-4.659(1-2x)^{2}+15.52(1-2x)^{4}
)\varphi_{as}(x)}
\end{equation}

As mentioned in section~\ref{the spacelike form factor}, the
two last distributions have slow logarithmic evolution with $Q$.
We ignore this evolution because it is quite
insignificant in the range of energy we consider here.

Following~\cite{hye---93,jakkro93}, one may
also include some intrinsic transverse behavior. We tried the
different forms of wave functions described in~\cite{jakkro93} and
found only small differences for the behavior of the ratio.
Therefore, we only quote here the sample form for
which we will show some results in the following:
\begin{equation}
\label{psi-transverse}
\hat{\psi}^{(0)}(x,b)=\varphi(x)e^{-b^{2}/4b^{2}_{0}}
\end{equation}
which is a simple way to modelize the transverse behavior without
a long set of parameters. \(2b_{0}\) related to the valence state
radius is proposed in~\cite{jakkro93}:
\(b^{2}_{0}=4.082\hskip 0.265em GeV^{-2}\).

Figs~\ref{transv}{--}\ref{distr} show our numerical results for
the meson form
factors in the large but non asymptotic \(Q^{2}=|q^{2}|\)
region~\footnote{In this sub-section, as there
is no confusion, we do not distinguish the absolute value of
spacelike and timelike transfers.}. In Fig.~\ref{transv},
\(Q^{2}|F_{\pi}(Q^{2})|\) is plotted against \(Q^{2}\) for both
timelike and spacelike regions up to \(Q=50\Lambda\). The
distribution considered is the asymptotic one,
Eq.~(\ref{phi-as}). The slow convergence of the timelike and
spacelike quantities is manifest while the counting rule
(\(F_{\pi}\propto 1/Q^{2}\)) is reasonably well describing the
$Q^{2}$ dependence down to a few $GeV^{2}$ in both cases. The
inclusion of the intrinsic $b$-dependence given by
Eq.~(\ref{psi-transverse}) (dashed
lines) does not significantly modify the results.

In Fig.~\ref{distr}(a), the modulus of the timelike form factor
(multiplied by
$Q^{2}$) is shown for the three choices of distribution amplitudes:
CZ form (solid line), asymptotic one (dashed line) and FHZ one
(long-dashed line). The experimental data shown comes from $\Psi$
decay~\cite{kahmil93}. Sudakov suppression has been
included but no intrinsic $b$-dependence. Fig.~\ref{distr}(b)
shows the ratio
of the timelike to the spacelike form factors. This ratio is rather
wave function independent and decreases very slowly to $1$ from a
value of around $2$ in most of the experimentally accessible range.

Although this ratio turns out to be quite difficult to reliably
extract from experimental data in the meson case, it is quite
straightforwardly measured in the proton case. We will analyse the
proton case in a forthcoming work. If we restrict to a simple
quark-diquark picture, we would get a timelike to spacelike ratio
quite similar to the one obtained here for the meson case, and thus
understand the experimental value.
\pagebreak

\section{Conclusion}

\hspace{\parindent}
In this paper, we showed that the difference between spacelike and
timelike form factors at large accessible transfer is predictible
from an improved perturbative QCD analysis. We understand at least
qualitatively the enhancement of the timelike values at large but
non asymptotic transfers as mostly due to the integrable
singularities of gluon and quark propagators. This strengthens the
faith in the applicability of perturbative reasoning at
intermediate energies (above a few GeV) at least for
semi-quantitative understanding of the strong interaction physics.

For the pion case, the uncertainties in the extraction of the
spacelike form factor~\cite{carmil90} show the need for another way
to access this observable, the simplest one in exclusive scattering.
We demonstrated that the usual formalism of Brodsky and Lepage has
to be improved to account for the differences
between timelike and spacelike regions in the energy range
experimentally reachable. More experimental data are still needed
to test our knowledge of the pion wave function.

The proton case is more interesting since the extraction of the
spacelike form factor is without ambiguity. The comparison of
spacelike and timelike form factors thus
appears to be a good way to understand the hadronic wave function.

Many other hard exclusive processes dwell on timelike transfers.
The \(\gamma \gamma \rightarrow \pi \pi , \hskip 0.265em  p\bar{p}\)
reactions
at fixed angle for instance demand a more careful analysis than
available now, not to speak of the difficult instances where
pinch singular diagrams mix up, as in the ratio of \(p \bar{p}\)
to $pp$ elastic scattering. More work needs to be done and
experimentally tested before we know for sure that exclusive
timelike reactions help us to understand confinement dynamics
through the unraveling of hadron wave functions in their lowest
Fock state.


\vspace{1cm}
{\it Acknowledgments.}
We thank P. Guichon, R. Jakob, P. Kroll, A. Mueller and J. Ralston
for useful discussions. CPT is Unit\'e Propre 014 du Centre
National de la Recherche Scientifique.

\newpage

\appendix

\section*{Appendix A}

\hspace{\parindent}
We give in this Appendix some arguments for the power suppression
of the oscillating factor we get in
sub-section~\ref{asymptotic behavior}.
Again we will concentrate on the simple case one has when
neglecting the transverse momentum in the quark propagator
(see sub-section~\ref
{asymptotic behavior}) and consider the integral analogous to
$I$ in Eq.~(\ref{asymptotic}):
\begin{equation}
I''=\int _{0}^{B}bdbK_{0}(\sqrt{xy}zb)
={{1}\over{xyz^{2}}}-{{B}\over{\sqrt{xy}z}} K_{1}(\sqrt{xy}zB).
\end{equation}

As in section \ref{the timelike form factor}, we should worry
about the behavior of the Bessel function with large argument (
$K_{1}$ has the same asymptotic behavior as $K_{0}$). We get:
\begin{equation}
I'' \approx {{1}\over{xyz^{2}}}-{{\sqrt{\pi B/2}}
\over{{(\sqrt{xy}z)}^{3/2}}}e^{-\sqrt{xy}zB}.
\end{equation}

In this explicit asymptotic form, one may guess that because the
phase varies rapidly with $x$ or $y$ due to the presence of the
large $|z|$ factor there may be some destructive interferences
when integrating the second term of $I''$. The presence of any
smooth distribution as weight functions will not destroy this
feature. To see this explicitely we can
perform the integration over some finite range for $x$ to avoid
region where the asymptotic form fails and
also to allow further simplifications. Precisely, we look at:
\begin{equation}
J=\int _{1/2-a}^{1/2+a}dx \varphi (x)I''(x,y={{1}\over{2}})
\end{equation}
and replaces $x=1/2$ everywhere in the integrand except in the
phase where we take the expansion of the square root of $x$
around $1/2$ up to first order. With these simplifications, we
easily get:
\begin{equation}
J={{8a}\over{z^{2}}} \varphi ({{1}\over{2}})
\left[ 1-\sqrt{{{\pi}\over{zB}}}
\hskip 0.265em e^{-{{zB}\over{2}}}
{{\sinh {{zBa}\over{2}}} \over{a}} \right],
\end{equation}
which with the replacement $z=-iW$ has a modified behavior
compared to $I''$ and the leading behavior is now identical in
the spacelike and timelike directions of the complex plane. However,
the previous equation, even if approximatively, still indicates
qualitatively that this identical asymptotic behavior may be reached
rather slowly.

\section*{Appendix B}

\hspace{\parindent}
We discuss in this Appendix the analytic continuation of the
Sudakov suppression factor.

In the complete expression for the form factor
Eq.~(\ref{F expression}), two factors: $T_{H}$ and $e^{-S}$, are
scale dependent and must {\it a priori} be analytically continued.
In sub-section~
\ref{asymptotic behavior}, we give some arguments to show that the
Sudakov factor has a leading behavior which is not affected by
analytic continuation so that we can study the contribution to
timelike form factor ignoring this kind of difficulties.
However, in the range of transfers we consider in the numerical
study of sub-section~
\ref{comparison}, these arguments may not apply. In $e^{-S}$, the
scale always appears in logarithms and after analytic continuation,
one gets a phase which is sub-leading compared to the remaining
large logarithm. The Sudakov exponent is known~\cite{botste89}
up to next-to-leading logarithms and we may keep the imaginary part
which comes from the leading-log (Eq.~(\ref{s-sSL})) as a
next-to-leading component. In this kind of analysis, the additional
real part due to the product of logarithms (the $\pi ^{2}$-factor),
which may lead to an aditional enhancement~\cite{magste90}, is
automatically dropped.
A further analysis of the amount of correction which may be provided
by such terms gives an \(O{(10\%)}\) extra enhancement of the
timelike form factor.

The effect of the imaginary part in the Sudakov exponent appears to
be more important. The ratio of the timelike to spacelike factor is:
\begin{eqnarray}
{{e^{-S_{TL}}}\over{e^{-S_{SL}}}}=
e^{i( \phi (x)+ \phi (1-x)+(x \leftrightarrow y))}, \nonumber
\\
\phi(x,W,b) \approx -{{C_{F} \pi} \over{4 \beta}}
\ln \hskip 0.265em {{- \ln \hskip 0.265em b}\over{\ln \hskip 0.265em
{{xW}\over{\sqrt{2}}}}}.
\label{phase}
\end{eqnarray}

We must define a prescription to cut-off the small $b$ region.
For the spacelike Sudakov suppression, Li and
Sterman~\cite{li-ste92} choose to include the exponential factor
only in the region of large $b$ defined by \(bxW>\sqrt{2}\) and
furthermore only when the sum of all exponents is negative and leads
effectively to a suppression. The first prescription is associated
with approximations discussed in
section~\ref{the spacelike form factor} and we will not questioned it
in the following. In the spacelike case, it appears that the second
constraint may be easily relaxed in a numerical study, because only a
very small enhancement results when forgetting this constraint.
We have observed that this is not likely to be the case for the
timelike form factor.

A numerical study, in the range of transfer $10$--$30 \Lambda$,
with the consideration of the total gluon propagator alone and the
prescription: $\phi$ from Eq.~(\ref{phase}) if $\rm{Re}(-S)<0$,
shows, for the asymptotic distribution, a $5$--$10\%$ depletion of
the timelike form factor with respect to the value without
consideration of phase, whereas for the CZ distribution, the
diminution is $20$--$30\%$.

\newpage


\newpage
\begin{figure}
\caption{The pion form factor.
\label{sl-ff}}
\end{figure}

\begin{figure}
\caption{Factorization of the form factor.
\label{conv}}
\end{figure}

\begin{figure}
\caption{Tree graph for $T_{H}$.
\label{tree-graph}}
\end{figure}

\begin{figure}
\caption{Leading radiative corrections grouped in the wave function.
\label{1-loop}}
\end{figure}

\begin{figure}
\caption{The timelike hard amplitude.
\label{tl-ff}}
\end{figure}

\begin{figure}
\caption{The real part (a) and modulus (b) of the deviation to
spacelike scaling $\Delta$. Solid (dashed) line is with
(without) the Sudakov suppression factor. W is in $\Lambda$ units.
\label{as}}
\end{figure}

\begin{figure}
\caption{Timelike and spacelike form factors: different
transverse behaviors. The x-dependence is the asymptotic form
of Eq.~(\protect\ref{phi-as}). Energies are in $\Lambda$ units.
\label{transv}}
\end{figure}

\begin{figure}
\caption{Timelike form factor (a) and timelike over spacelike
ratio (b): sensitivity to
distribution. CZ is Eq.~(\protect\ref{phi-CZ}),
FHZ~(\protect\ref{phi-FHZ}), as~(\protect\ref{phi-as});
experimental data from~\protect\cite{kahmil93}.
\label{distr}}
\end{figure}

\end{document}